# Integrating the Fermi Gamma-Ray Burst Monitor into the 3rd Interplanetary Network

K. Hurley

*UC Berkeley Space Sciences Laboratory, Berkeley, CA 94708*

M. Briggs, V. Connaughton

*University of Alabama, Huntsville, AL 35805*

C. Meegan

*USRA Huntsville*

T. Cline

*Emeritus, NASA Goddard Space Flight Center, Greenbelt MD 20771*

I. Mitrofanov, D. Golovin, M. L. Litvak, A. B. Sanin

*Institute for Space Research, Profsojuznaja 84/32, Moscow 117997, Russian Federation*

W. Boynton, C. Fellows, K. Harshman, R. Starr

*University of Arizona, Lunar and Planetary Laboratory, Tucson, AZ 85721*

S. Golenetskii, R. Aptekar, E. Mazets, V. Pal'shin, D. Frederiks

*Ioffe Physico-Technical Institute of the Russian Academy of Sciences, St. Petersburg, 194021, Russian Federation*

D. M. Smith

*Department of Physics and Santa Cruz Institute for Particle Physics, University of California, Santa Cruz, 1156 High Street, Santa Cruz, CA 95064*

C. Wigger

*Paul Scherrer Institute, 5232 Villigen PSI, Switzerland*

A. Rau, A. von Kienlin

*Max-Planck-Institut für extraterrestrische Physik, Giessenbachstrasse, Garching, 85748 Germany*

K. Yamaoka

*Department of Physics and Mathematics, Aoyama Gakuin University, 5-10-1 Fuchinobe, Sagamihara, Kanagawa, 229-8558, Japan*

M. Ohno

*Institute of Space and Astronautical Science (ISAS/JAXA), 3-1-1 Yoshinodai, Sagamihara, Kanagawa 229-8510, Japan*

Y. Fukazawa

*Department of Physics, Hiroshima University, 1-3-1 Kagamiyama, Higashi-Hiroshima, Hiroshima 739-8526, Japan*

T. Takahashi

*Institute of Space and Astronautical Science (ISAS/JAXA), 3-1-1 Yoshinodai, Sagamihara, Kanagawa 229-8510, Japan*

M. Tashiro, Y. Terada

*Department of Physics, Saitama University, 255 Shimo-Okubo, Sakura-ku, Saitama-shi, Saitama 338-8570, Japan*

T. Murakami

*Department of Physics, Kanazawa University, Kadoma-cho, Kanazawa, Ishikawa 920-1192, Japan*

K. Makishima

*Department of Physics, University of Tokyo, 7-3-1 Hongo, Bunkyo-ku, Tokyo 113-0033, Japan*

S. Barthelmy, J. Cummings, N. Gehrels, H. Krimm

*NASA Goddard Space Flight Center, Greenbelt MD 20771*

J. Goldsten

*Applied Physics Laboratory, Johns Hopkins University, Laurel, MD 20723*

E. Del Monte, M. Feroci
*IASF/INAF, 00133 Rome, Italy*
M. Marisaldi
*INAF/IASF, 40129 Bologna, Italy*

We are integrating the Fermi Gamma-Ray Burst Monitor (GBM) into the Interplanetary Network (IPN) of Gamma-Ray Burst (GRB) detectors. With the GBM, the IPN will comprise 9 experiments. This will 1) assist the Fermi team in understanding and reducing their systematic localization uncertainties, 2) reduce the sizes of the GBM and Large Area Telescope (LAT) error circles by 1-4 orders of magnitude, 3) facilitate the identification of GRB sources with objects found by ground– and space-based observatories at other wavelengths, from the radio to very high energy gamma-rays, 4) reduce the uncertainties in associating some LAT detections of high energy photons with GBM bursts, and 5) facilitate searches for non-electromagnetic GRB counterparts, particularly neutrinos and gravitational radiation. We present examples and demonstrate the synergy between Fermi and the IPN. This is a Fermi Cycle 2 Guest Investigator project.

## 1. INTRODUCTION

We are using the data from the Fermi Gamma-ray Burst Monitor (the GBM, described elsewhere in this meeting), in conjunction with the data from 8 other missions in the Interplanetary Network (IPN), to derive the positions of gamma-ray bursts by triangulation. The IPN is an all-sky, full-time monitor of transient activity, when the integrated response of all the instruments in the network is considered, and has a limiting accuracy of <1'. Its current event detection rate is ~325/year above a fluence threshold of ~$10^{-6}$ erg cm$^{-2}$, or a peak flux threshold of 1 photon cm$^{-2}$ s$^{-1}$, considering only those bursts detected by two or more spacecraft. This makes it possible to study a wide variety of events which narrow field-of-view imaging GRB instruments like the SuperAGILE, INTEGRAL-IBIS, and Swift BAT will seldom detect. These include very intense bursts, very long bursts, repeating sources (gravitationally lensed GRBs and bursting pulsars like GRO1744-28 are two examples), soft gamma repeater activity, and possibly other as-yet undiscovered phenomena.

## 2. THE IPN

The IPN comprises 8 missions, besides Fermi. Four of them (AGILE, RHESSI, Suzaku, and Swift), are in low Earth orbit. INTEGRAL is in an eccentric orbit with an apogee of about 0.5 light-seconds, and Wind is at distances up to about 7 light-seconds from Earth. The two distant missions are Mars Odyssey, in orbit around Mars, and MESSENGER, on its way to Mercury, at distances up to 1000 and 600 light seconds, respectively. Depending on how many spacecraft detect a burst, and what their distances are, triangulation gives localization regions which can be very small error boxes or ellipses with areas of a few square arcminutes, large error boxes with areas of many square degrees, or annuli.

## 3. BURST STATISTICS

We have compiled statistics of IPN detections from the launch of Fermi through May 2009. The GBM detected 195 bursts in that period (one every 1.7 days). Table 1 shows how many of those bursts were detected by 1, 2, …, 9 IPN spacecraft, considering only those bursts with a GBM detection. Table 2 shows the number of events detected by each IPN spacecraft.

Table 1: The number of bursts detected by N spacecraft, where N=1,2,…9.

| Number of Spacecraft | Number of Bursts |
|---|---|
| 1 | 32 |
| 2 | 44 |
| 3 | 33 |
| 4 | 31 |
| 5 | 22 |
| 6 | 18 |
| 7 | 13 |
| 8 | 1 |
| 9 | 1 |

Table 2. The total number of bursts observed by each of the IPN spacecraft. The number of bursts outside the coded fields of view of SuperAGILE, INTEGRAL-IBIS, and Swift BAT are given in parentheses

| Spacecraft | Number of Bursts |
|---|---|
| AGILE | 24 (22) |
| Fermi | 195 |
| INTEGRAL | 92 (90) |
| Konus | 108 |
| Mars Odyssey | 20 |
| MESSENGER | 52 |
| RHESSI | 39 |
| Swift | 64 (29) |
| Suzaku | 75 |

## 4. SOME RESULTS TO DATE

Figures 1-11 illustrate some of the results obtained so far. The GBM statistical-only contours are shown, as well as the error circles, which include 3° systematic errors, and the best position (asterisk); where shown, the LAT error circles include 0.1° systematic uncertainties. Many of the IPN localizations are still preliminary, and can be improved substantially; for example, error ellipses can be derived instead of annuli.

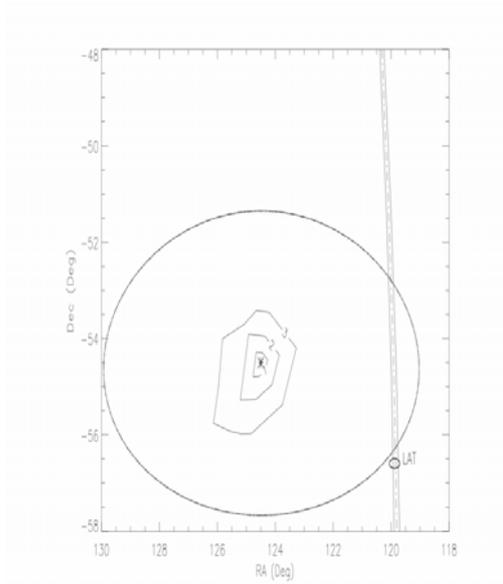

Figure 1. GRB080916, observed by the LAT and the GBM, as well as by Konus, RHESSI, INTEGRAL SPI-ACS, MESSENGER, and AGILE. Only the Konus-MESSENGER annulus is shown; it is consistent with the position of the optical afterglow (not shown).

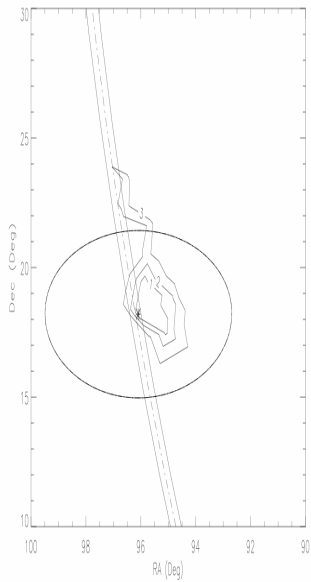

Figure 2. GRB080925. This burst was observed by the GBM, but not the LAT. In addition, it was detected by Konus, RHESSI, MESSENGER, and Suzaku. Only the Konus-MESSENGER annulus is shown; it reduces the size of the error circle by a factor of about 11.

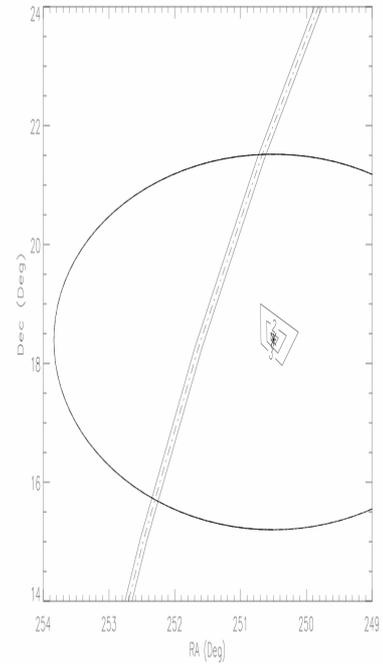

Figure 3. GRB081009. This burst was observed by the GBM, but not the LAT. It was also observed by Konus, RHESSI, INTEGRAL SPI-ACS, Swift (but outside the coded field of view), and MESSENGER. The Konus-MESSENGER annulus is shown; it defines an area which is about 50 times smaller than the GBM error circle.

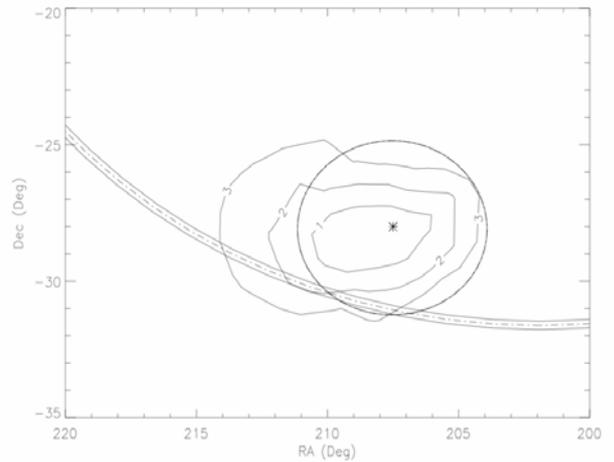

Figure 4. GRB081101. This is a Konus/INTEGRAL SPI-ACS/Swift/MESSENGER/GBM burst, which was outside the coded field of view of the BAT. The Konus-MESSENGER annulus is shown; it reduces the area of the GBM circle by a factor of about 20.

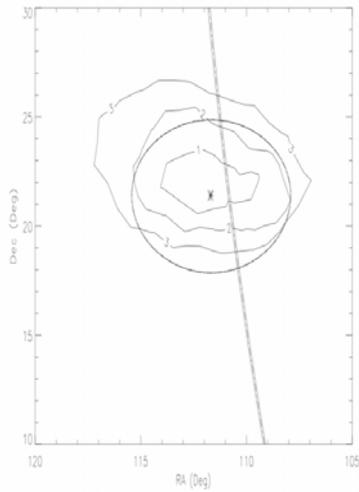

Figure 5. GRB081110. A Konus/INTEGRAL SPI-ACS/Swift/MESSENGER/AGILE/GBM event, outside the coded field of view of Swift. The Konus-MESSENGER annulus is shown.

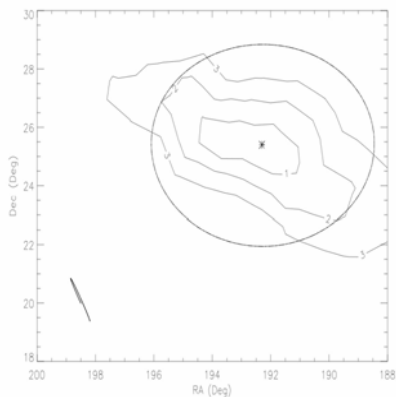

Figure 6. GRB090112, observed by the GBM, Mars Odyssey, Konus, RHESSI, Swift, MESSENGER, and Suzaku. The burst was outside the coded field of view of the Swift BAT; the IPN error ellipse has an area of 280 square arcminutes.

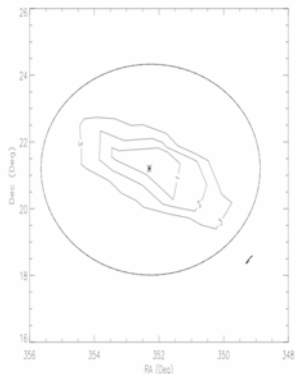

Figure 7. GRB090131, a Mars Odyssey/Konus/RHESSI/INTEGRAL SPI-ACS/Swift/MESSENGER/Suzaku/AGILE event, which was outside the Swift coded field of view. The IPN error ellipse has an area of 27 square arcminutes.

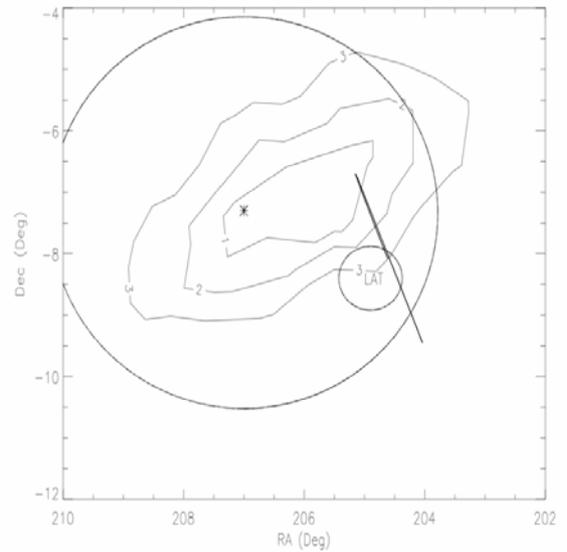

Figure 8. GRB090217. This burst was observed by the GBM and the LAT, and Swift did follow-up observations. The XRT found 3 sources (S1, S2, and S3, not shown), of which one (S1) lies within the 230 square arcminute IPN error ellipse.

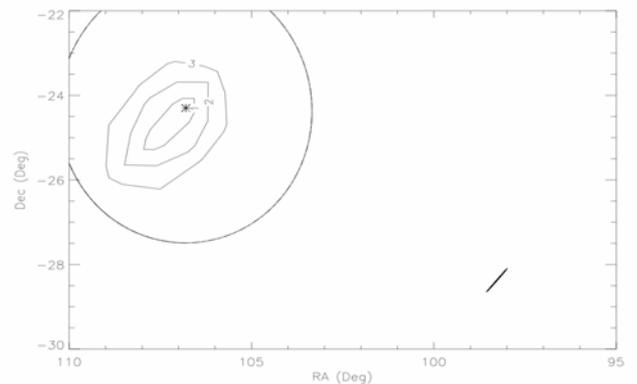

Figure 9. GRB090228, observed by the GBM, Mars Odyssey, Konus, RHESSI, MESSENGER, and AGILE. The IPN error ellipse area is 34 square arcminutes. The source of the discrepancy is believed to lie in the estimated systematic uncertainty of the GBM localization.

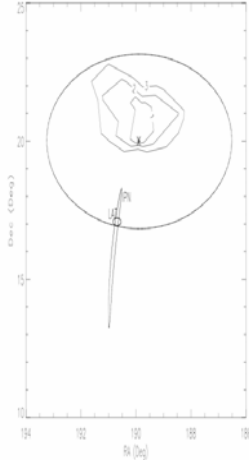

Figure 10. GRB090323. This burst was observed by the LAT and the GBM, and an optical counterpart was identified for it. It was also observed by Mars Odyssey, Konus, INTEGRAL SPI ACS, Swift (outside the coded field of view), and MESSENGER. The IPN error ellipse has an area of 1870 square arcminutes, and is consistent with the optical counterpart (not shown).

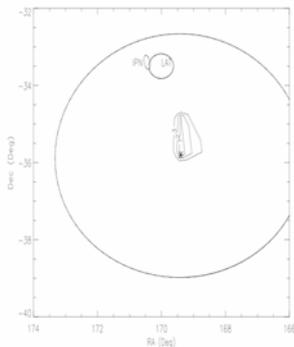

Figure 11. GRB090626, observed by the GBM, the LAT, Mars Odyssey, Konus, RHESSI, INTEGRAL SPI ACS, Swift (outside the coded field of view), MESSENGER, and Suzaku. The IPN error ellipse has an area of 160 square arcminutes.

## 5. CONCLUSIONS

The IPN localizes many GBM bursts for which no other precise localization exists, and for these events, the IPN localizations can be used to reduce the error box areas significantly. This is also true even for some LAT bursts. The GBM has the advantage of excellent statistics and time-tagged data, which makes it a very valuable addition to the IPN. Even in those cases where a burst is observed only by the GBM and Konus, and no distant spacecraft, the resulting IPN annulus often reduces the GBM error box area.

IPN bursts, regardless of their error box sizes or the delays in producing them, have been used for a wide variety of studies, such as searching for VHE gamma rays with ground-based instrumentation, for neutrinos, and for gravitational radiation. They have also been used for searching for coincidences between GRBs and Type Ib/c supernovae. These studies benefit from a large database, and they tend to target the GRBs which are more intense and, on average, closer to Earth. Thus the IPN database is a good source of events for them.

The IPN localizations are public. For more details, refer to the IPN website: ssl.berkeley.edu/ipn3/index.html, or contact khurley@ssl.berkeley.edu.

### Acknowledgments

KH is grateful for IPN support from the NASA Guest Investigator programs for Fermi (NNX09AU03G), Swift (NNX09AO97G), INTEGRAL (NNX09AR28G), and Suzaku (NNX09AV61G), and for support under the Mars Odyssey and MESSENGER Participating Scientist Programs, JPL Contract 1282046 and NASA NNX07AR71G, respectively. The Konus-Wind experiment is supported by a Russian Space Agency contract and RFBR grant 09-02-00166a.